\documentclass[reprint,
latex apssamp.tex,
bibtex apssamp
bibnotes,
 amsmath,amssymb,
 aps,
pra, superscriptaddress]{revtex4-2}

\newcommand{\be}{\begin{equation}}
\newcommand{\ee}{\end{equation}}
\newcommand{\bea}{\begin{eqnarray}}
\newcommand{\eea}{\end{eqnarray}}
\usepackage{amssymb,amsmath,amsthm}
\usepackage{xcolor}
\usepackage{graphicx}
\usepackage{dcolumn}
\usepackage{bm}
\usepackage{physics}
\usepackage{braket}
\usepackage{qcircuit}
\usepackage[english]{babel}
\newtheorem{theorem}{Proposition}
\newtheorem{lemma}{Lemma}
\newtheorem{claim}{Claim}
\newtheorem{definition}{Definition}

\usepackage{url}

\DeclareFixedFont{\ttb}{T1}{txtt}{bx}{n}{10}
\DeclareFixedFont{\ttm}{T1}{txtt}{m}{n}{10}

\usepackage{color}
\definecolor{deepblue}{rgb}{0,0,0.5}     
\definecolor{deepred}{rgb}{0.6,0,0}
\definecolor{deepgreen}{rgb}{0,0.5,0}

\newcommand{\blue}[1]{{\color{black}#1}}

\usepackage{listings}

\newcommand\pythonstyle{\lstset{
		language=Python,
		basicstyle=\ttm,
		keywordstyle=\ttb\color{deepblue},
		breaklines=true,
		breakatwhitespace=false,
		emphstyle=\ttb\color{deepred},
		stringstyle=\color{deepgreen},
		frame=tb,                         
		showstringspaces=false
}}

\usepackage[colorlinks=true,pdfborder={0 0 0},linkcolor=blue]{hyperref}

\begin{document}

\title{
Multipartite entanglement and quantum error identification in $D$-dimensional cluster states}
\author{Sowrabh Sudevan}
\email{ss18ip003@iiserkol.ac.in}
\affiliation{Indian Institute of Science Education and Research Kolkata
\\Mohanpur, Nadia-741 246, West Bengal, India}
\author{Daniel Azses}
\email{danielazses@mail.tau.ac.il}
\affiliation{School of Physics and Astronomy, Tel Aviv University, Tel Aviv 6997801, Israel}
\author{Emanuele G. Dalla Torre}
\email{emanuele.dalla-torre@biu.ac.il}
\affiliation{Department of Physics, Bar-Ilan University, 52900 Ramat Gan, Israel}
\author{Eran Sela}
\affiliation{School of Physics and Astronomy, Tel Aviv University, Tel Aviv 6997801, Israel}
\author{Sourin Das}
\email{sourin@iiserkol.ac.in}
\affiliation{Indian Institute of Science Education and Research Kolkata
\\Mohanpur, Nadia-741 246, West Bengal, India}

\begin{abstract}
An entangled state is said to be $m$-uniform if the reduced density matrix of any $m$ qubits is maximally mixed. This is intimately linked to pure quantum error correction codes (QECCs), which allow not only to correct errors, but also to identify their precise nature and location. Here, we show how to create $m$-uniform states using local gates or interactions and elucidate several QECC applications. We first show that $D$-dimensional cluster states are $m$-uniform with $m=2D$. \blue{This zero-correlation length cluster state does not have finite size corrections to its $m=2D$ uniformity, which is exact both for infinite and for large enough but finite lattices. Yet at some finite value of the lattice extension in each of the $D$ dimensions, which we bound, the uniformity is degraded due to finite support operators which wind around the system. We also outline how to achieve larger $m$ values using quasi-$D$ dimensional cluster states.} This opens the possibility to use cluster states to benchmark errors on quantum computers. We demonstrate this ability on a superconducting quantum computer, focusing on the 1D cluster state which, we show, allows to detect and identify 1-qubit errors, distinguishing $X$, $Y$ and $Z$ errors.
\end{abstract}

\maketitle

\section{\label{sec:level1}Introduction}
Bipartite entanglement is a well understood concept, as it can be clearly quantified by the appropriate reduced density matrix \cite{sakurai_napolitano_2017,nielsen_chuang_2010}. Quantifying multipartite entanglement is much more challenging and was studied in the literature using many distinct measures \cite{Vedral2008,bengtsson_zyczkowski_2006,HORODECKI1994145,Meyer_2002,https://doi.org/10.48550/arxiv.1612.07747,Horodecki_2009,Enríquez2016maximally,Hein2004Multiparty,PhysRevA.64.022306,Markham_2007,schatzki2022hierarchy}. A common approach to multipartite entanglement relies on the entanglement across all possible bipartitions. This approach was first introduced by Ref.~\cite{page1993average} in a study of the average entanglement of random pure states. Later studies led to the definition of $m$-uniformity: multi-qubit states in which all the reduced density matrices of $m$ qubits are maximally mixed \cite{calderbank1998quantum,scott2004multipartite,Facchi_2008,arnaud2013exploring,shi2006construction,goyeneche2014genuinely,PhysRevA.104.032601,raissi2020constructions}.

A simple example of $m$-uniformity is given by the $n$-qubit GHZ state, $\ket{\psi}_{\rm GHZ} = (\ket{0}^{\otimes n} + \ket{1}^{\otimes n})/\sqrt{2}$. Any 1-qubit subsystem $A_1$ corresponds to the reduced density matrix $\rho_{A_1} = \left( \ket{0}\bra{0} + \ket{1}\bra{1} \right)/2$, which is maximally mixed, i.e. proportional to the identity operator in the subsystem. Hence, the GHZ state is at least 1-uniform. For a subsystem $A_2$ of any two qubits, one has $\rho_{A_2} = \left( \ket{00}\bra{00} + \ket{11} \bra{11} \right)/2$ which is not maximally mixed. Therefore, the  GHZ state is only $1$-uniform~\cite{goyeneche2014genuinely,ruizgonzalez2022digital}.

The notion of $m$-uniformity has deep links to quantum error correction codes (QECC). In the case of quantum states that encode no logical information, Ref.~\cite{scott2004multipartite} proved that any $m$-uniform state has the ability to locate and identify a quantum error, assuming that it acted at most on $ \lfloor m/2\rfloor$ qubits \cite{calderbank1998quantum}, or equivalently, a sequence of at most $ \lfloor m/2\rfloor$  single-qubits errors.  The relationship between $m$-uniformity and QECC was extended to states that encode a finite amount of quantum information~\cite{huber2020quantum}, \blue{and is related to quantum information scrambling~\cite{kelly2022coherence,weinstein2022scrambling}}. Note that $m$-uniformity is a sufficient but not necessary condition for error correction: A known example is Kitaev's toric code, which is only $3$-uniform, but can correct an extensive number of errors \cite{kitaev1997fault}. This code is an example of a non-pure QECC, which can correct errors even without being able to fully identify them. In contrast, $m$-uniform states give rise to pure, i.e. non-degenerate QECCs, where the errors are first fully identified and only then corrected \cite{nielsen_chuang_2010}. Hence, pure codes are particularly useful in benchmarking noisy quantum computers.

Creating states with large $m$-uniformity is a key challenge in quantum information. Earlier studies discussed how to perform this task using orthogonal arrays \cite{shi2006construction,goyeneche2014genuinely,pang2019two,pang2021quantum}, numerical methods \cite{borras2007multiqubit,Facchi_2008}, graph states \cite{helwig2013absolutely,raissi2022general,sudevan2022nqubit} and other constructions \cite{raissi2018optimal,zang2021quantum,raissi2020modifying,raissi2020constructions}. Special attention was drawn to $n$-qubit states that are $\lfloor n/2 \rfloor$-uniform, also known as absolutely maximal entangled (AME), which have important applications in quantum secret sharing and quantum teleportation~\cite{helwig2012absolute,raissi2018optimal}. Their existence was proven for $n=2,3,5,6$ qubits only \cite{scott2004multipartite,huber2017absolutely}. Moreover, using theoretical enumerator tools from QECC, useful bounds on their existence for qudits were derived \cite{scott2004multipartite,huber2018bounds}. In general, it is always possible to construct $m$-uniform states for a desired $m$ if the number of qubits is large enough \cite{goyeneche2014genuinely,arnaud2013exploring}, but determining the minimal number of qubits required for a given $m$ is, as far as we know, still an open question though some lower bounds are known~\cite{rains1999quantum,arnaud2013exploring,huber2020quantum}. 
In addition, these constructed states may be highly non-local and pose a challenge to prepare them on a quantum computer using low depth circuits. 

In this paper we address this challenge by focusing on cluster states, which can be realized as ground states of cluster Hamiltonians, \blue{and are sometimes referred to as} graph states~\cite{PhysRevLett.86.910}. Cluster Hamiltonians can be expressed as a sum of local terms, known as stabilizers, and are commonly used to describe condensed matter systems. As we will explain, the degree of uniformity of these states are bounded by the support of the stabilizers, i.e. the size of the neighborhood that interacts with each qubit. The ground state of these Hamiltonians can be prepared exactly on quantum computers via shallow circuits of local unitary gates \cite{choo2018measurement,azses2020identification,smith2022crossing}. Cluster states in one and two dimensions have been studied in the context of symmetry protected topological states of matter~\cite{smacchia2011statistical,son2011quantum,azses2020identification}, and can be used in measurement based quantum computation and teleportation~\cite{briegel2009measurement,raussendorf2019computationally,NIELSEN2006147,azses2020identification,lee2022measurement}. 

We show that the cluster state in $D$-dimensions is $2D$-uniform. \blue{When considering the cluster state as a stabilizer QECC it has no logical qubits encoded}. Despite of not containing a logical space, this state allows for quantum error
detection of $D$ individual errors. This ability of error detection stems from a connection between $m$-uniformity and QECCs that allows us to employ these states for benchmarking errors that may act on up to $ \lfloor m/2 \rfloor $ qubits in current noisy-intermediate scale quantum computers~\cite{preskill2018quantumcomputingin}. We demonstrate this procedure on a real quantum computer for $D = 1$. Finally, we provide examples of $m-$uniform subspaces in which a finite logical space is encoded.

The paper is organized as follows. In Sec.~\ref{sec:2} we introduce the cluster states and determine their $m-$uniformity based on the stabilizer formalism. In Sec.~\ref{sec:3} we review the connection to QECC and deduce the quantum error detection ability of cluster states. We present a demonstration of this application in Sec.~\ref{sec:4}. In Sec.~\ref{sec:5} we exemplify how $m$-uniform spaces, encoding logical information and also allowing to detect errors, can be constructed. We conclude in Sec.~\ref{sec:6}.

\begin{figure}
\centering
\includegraphics[width=\linewidth]{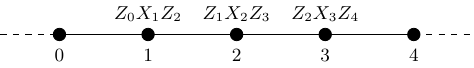}
\caption{
\blue{One dimensional cluster chain depicted using graphical notation of vertices and lines. The vertices represent the qubits, which are all prepared in the $\ket{+}$ state. The lines connecting two qubits represent a controlled-Z gate. We notate a few commuting stabilizer operators appearing in Eq.~(\ref{cluster}) on their respective qubits.}}
\label{fig:1D_Cluster_Ising} 
\end{figure}

\section{Cluster states and their multipartite entanglement}
\label{sec:2}
The $D$-dimensional cluster Hamiltonian describes $n$ qubits located on the vertices of a $D$-dimensional square lattice \blue{with axes lengths $L_1,L_2,\dots,L_D$, where $n=L_1 \cdot  L_2 \cdots L_D $}, such that each qubit interacts with exactly $2D$ neighbours. Unless specified otherwise, we consider \blue{a finite system with} periodic boundary conditions (PBC) in each dimension. The one dimensional ($D=1$) cluster  Hamiltonian, which is depicted in Fig.~\ref{fig:1D_Cluster_Ising}, is defined as
\begin{equation}
\label{cluster}
    H = -\sum_i Z_{i-1}X_{i}Z_{i+1}.
\end{equation}
The minus sign is motivated by the ferromagnetic ground state of the Ising model and its generalizations to the cluster Ising model~\cite{guo2022emergent, ding2019phase, nie2017scaling, giampaolo2015topological, giampaolo2014genuine, smacchia2011statistical}. 
In $D>1$ dimensions, we denote the location of each qubit 
using a lattice vector $\mathbf{v}=(i_{1},i_{2},\dots,i_{D})$, \blue{where $i_k \in \mathbb{Z}$ and $-L_k/2 < i_{k} \leq L_k/2$ for each dimension $1 \le k\leq D$. 
The basis vectors $\{\mathbf{e}_{i}\}$ of the lattice
have 1 in their $i$'th entry and 0 elsewhere.}
The cluster Hamiltonian is then defined as
\be
\label{eq3}
    H=-\sum_{\mathbf{v}}X_{\mathbf{v}}{\Big(}\prod_{i=1}^D Z_{\mathbf{v}- \mathbf{e}_{i}} Z_{\mathbf{v}+ \mathbf{e}_{i}} {\Big)}
    =-\sum_{\mathbf{v}}s_{\mathbf{v}}.
\ee
Here, \blue{each vertex induces an operator $s_{\mathbf{v}}$. The $s_{\mathbf{v}}$'s are referred to as stabilizers, as they fulfill two special properties}~\cite{hein2006entanglement}: (i) they square to one, ${s_{\mathbf{v}}}^{2}=I$, because they correspond to tensor products of Pauli matrices; (ii) they commute, $[s_{\bf v},s_{\bf v'}]=0$. \blue{These properties simplify the problem of finding the groundstate.}

These two properties ensure that the Hamiltonian in Eq.~(\ref{eq3}) is frustration-free \cite{sattath2016when}, i.e. any ground state of $H$ is a simultaneous ground state of each  $s_{\mathbf{v}}$. In a more formal way, we note that the set of $q$ stabilizers $\big\{s_{\mathbf{v}}\big\}$ generates a $\mathbb{Z}_2^q$ group called the stabilizer group $\mathcal{S}$ \blue{by considering all their multiplications} \cite{gottesman1997stabilzier,zeng2015quantum}. \blue{To differentiate between the elements of $\mathcal{S}$ and their generators we notate the generators of the stabilizer group by $\{ s_i\}$ ($i=1,\dots, q$) and the elements of this group as $\mathcal{S}_i$ $(i=1, \dots, 2^{q})$}.
\blue{Focusing on the} cluster Hamiltonian with PBC, $q$ equals the number of qubits, $q=n$. Since the $\mathcal{S}_i$'s commute, they have a common set of eigenvectors and their eigenvalues are all $\pm 1$ as of the first property above. \blue{In this case, where the size of $\mathcal{S}$ equals the Hilbert space dimension one has a basis given by the eigenvectors}. Consider the unique eigenvector $|cs\rangle$ such that $s_i|cs\rangle=|cs\rangle$ for any $i$. By definition, \blue{it is the} unique ground state of the cluster Hamiltonian is the cluster state in $D$ dimensions and \blue{its ground state energy is} $-n$. For later reference, we define the {\it support} of $\mathcal{S}_i$ as the number of non-identity local Pauli matrices of $\mathcal{S}_i$. For example, for the 1-dimensional cluster state the support of each local term is ${\rm{supp}}(s_i)= 3$, while for the $D$-dimensional cluster state we have ${\rm{supp}}(s_i)= 2D+1$.

\subsection{\label{sec:level3}Reduced density matrices and stabilizer subgroups}
The stabilizer formalism leads to an explicit way to construct reduced density matrices. By definition, the pure density matrix $\rho$ of the cluster state corresponds to the projector into the $+1$ eigenvector of the generators of the stabilizer group \blue{$P_i = \frac{I+s_i}{2}$ and can
be written as \cite{hein2006entanglement}}
\begin{equation}
\label{eq:dm_s_formula}
    \rho \blue{= \prod_{i=1}^n P_i =} \frac{1}{2^{n}} \sum_{\sigma \in \mathcal{S}} \sigma.
\end{equation}
Using this expression, one can show that the reduced density matrix over the set of qubits $A$ is \cite{hein2006entanglement}
\begin{equation}
\label{eq:reduced_sa_factorization}
    \rho_A=\frac{1}{2^{|\mathcal{S}_{A}|}}\sum_{\sigma\in \mathcal{S}_A}\sigma,
\end{equation}
where $\mathcal{S}_A$ is the subgroup of the  stabilizer group $\mathcal{S}$ that has support {\it only} on the set of qubits $A$ and $|\mathcal{S}_{A}|$ is the number of elements in $\mathcal{S}_{A}$. For the subsystem $A$ to be maximally mixed, $\mathcal{S}_A$ should be the trivial group containing only the identity element $I$, or \begin{equation}
    \mathcal{S}_A = \big\{I\big\} \iff \rho_{A}\propto I.
\end{equation}
As mentioned, if this property applies to all sets of $m$ qubits, the state is defined to be $m$-uniform.

For example, consider the $3$-qubit cluster state in $D=1$ with PBC, i.e. the stabilizer state generated by $\{ s_i\} = \{X_1Z_2Z_3,Z_1 X_2 Z_3,Z_1Z_2 X_3\}$, which is the ground state of $H=-\sum_i s_i$. The eight elements of the stabilizer group $\mathcal{S}$ consist of $\{s_i\} \bigcup \{ I,  Y_1Y_2 I_3,Y_1 I_2Y_3,I_1 Y_2Y_3,-X_1X_2X_3 \}$. We can see that while the reduced density matrix over any single qubit is maximally mixed, the reduced density matrix over 2 qubits is not proportional to the identity. For example applying Eq.~(\ref{eq:reduced_sa_factorization}) for $A = \big\{1,3\big\}$ we have
\begin{equation}\label{example_S_A}
    \rho_{A}=\frac{1}{4}{\Big(}I+Y_1Y_3{\Big)},
\end{equation}
as $\mathcal{S}_A$ contains $\big\{I,Y_1Y_3\big\}$. Hence this state is 1-uniform but not 2-uniform.

\subsection{The infinite $D-$dimensional cluster state is $2D$-uniform}
\label{se:main_result}
\begin{theorem}
For a $D$-dimensional cluster state on an infinite lattice, if $|A|\leq 2D$ then $\mathcal{S}_A$ is the trivial subgroup, consisting only of the identity matrix acting in subsystem $A$, \blue{where $|A|$ is the number of qubits in subsystem $A$.}
\end{theorem}
This follows from the intuitive fact that the generators of the stabilizer group for the cluster states on a large enough lattice, having support $2D+1$, minimize the support of the  stabilizer group. We provide a proof of this proposition in Appendix~\ref{se:app_cs_2d_uniform}. \blue{To show its generality, we also consider different lattice structures in Appendix~\ref{app:lattice_structures}}. From this proposition, it immediately follows that \emph{the $D$-dimensional cluster state is $2D$-uniform on a large enough lattice}.

\blue{The cluster state does not have finite size corrections to its $m=2D$ uniformity, which is exact both for infinite and for large enough but finite lattices. Yet at a finite value of the lattice extension the uniformity is degraded due to finite support operators which wind around the system. The $4$-qubit $1D$ cluster state, which is not $2$-uniform, exemplifies this point. }

This leaves the question of what the lower bound for the system size in a given dimension $D$ is, in order to preserve the $2D-$uniformity. One can formally ask this question for a cluster state of $L^D$ qubits with PBC. In Appendix~\ref{se:proof_finite_lattice} we demonstrate that Proposition 1 in fact holds for any dimension, as long as $L \ge 8$. This proof does not saturate the lowest bound. In $1D$, 5 qubits with PBC are sufficient to obtain $2D$ uniformity \cite{scott2004multipartite}. By numerically inspecting the stabilizers and their multiplications for $D=1, 2, 3$, we  conjecture that the minimal size necessary for $2D$-uniformity is  $L = 5$ in each direction with PBC.

\subsection{Extended cluster states}
\begin{figure}[t] 
\centering
\includegraphics[width=\linewidth]{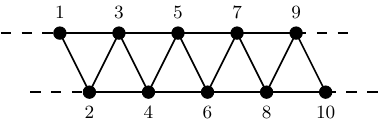}
\caption{Extended $ZZXZZ$ cluster state with next nearest neighbour interactions. \blue{In this case, the commuting stabilizers are supported on 5 sites, e.g. $Z_1 Z_2 X_3 Z_4 Z_5, Z_2 Z_3 X_4 Z_5 Z_6$.}}
\label{ladder}
\end{figure}
Can uniformity be increased by varying the range or support of stabilizers? For example, in 1D, rather than considering the support-3 $ZXZ$ operators in Eq.~(\ref{cluster}), one can consider the graph state defined by the stabilizers $s_i^{(p)} = Z_{i-p}\dots Z_{i-1} X_i Z_{i+1} \dots Z_{i+p}$. For $p=2$, this state can be created by applying controlled Z (CZ) gates on a ladder graph, as shown in Fig.~\ref{ladder}. While for $p=1$ we recover the cluster state which is $2$-uniform, it is not hard to see that all $p>1$ states on an infinite lattice are $3$-uniform. 

\blue{We provide a simple argument for the simple case of $p=2$, which can then be easily extended. The weight of a stabilizer generators sets an upper bound on the uniformity of a state, $m<5$. Is the state 4-uniform? To show that the uniformity is actually smaller, one should find operators of a support 4 or less which acquire a finite expectation. Consider the stabilizer group element $s_{i}s_{i+1} = Z_{i-p}\dots Z_{i-1} X_i Z_{i+1} \dots Z_{i+p} \cdot Z_{i-p+1}\dots Z_{i} X_{i+1} Z_{i+2} \dots Z_{i+p+1} = Z_{i-p}Y_i Y_{i+1} Z_{i+1+p}$. The weight of this element is $4$ for $p=2$. It is possible to check that there are no stabilizer group elements of a  smaller support. Hence the extended 1D cluster state for $p=2$ is at most $3$-uniform.}

The same result applies to stabilizers of the form $Z_{i} X_{i+1} \dots X_{i+p} Z_{i+p+1}$, which are related to a family of topological states~\cite{rajak2019complete}. An analogous construction can be used to create cluster states with larger $m$-uniformity in $D>1$ dimensions. 

\section{Cluster states as pure QECCs}
\label{sec:3}

In this section, we review basic definitions of QECCs and explain their connection to $m$-uniformity. 
Consider a logical subspace of dimension $2^k$, corresponding to $k$ logical qubits out of the Hilbert space of $n$ physical qubits, and denote its basis states by $\{ |i\rangle \}$. If, for a given positive integer $d$, the full set of operators, referred to as errors, $\{ E_a \}$ with ${\rm{supp}} (E_a) < d$, satisfy
\be
\label{eq:defEa}
\langle i | E_a | j \rangle=C(E_a) \delta_{ij},
\ee
then we say that the subspace is a QECC with distance $d$, and denote it with $[[n,k,d]]$ . Such QECC allows the correction of errors with support $\lfloor (d-1)/2 \rfloor$~\cite{nielsen_chuang_2010}.

For a general stabilizer code with a finite encoded subspace ($k>0$) the distance is given by $d={\rm{min}}[{\rm{supp}} \left( C(\mathcal{S}) - \mathcal{S}\right)]$, where $C(\mathcal{S})$ is the centralizer of the stabilizer group, i.e. the set of Pauli strings that commute with all the stabilizers. In other words, $d$ is the minimal support of operators that commute with $\mathcal{S}$ but not with the logical  operators. A stabilizer state without an encoded space $(k=0)$ is a  $[[n,0,d]]$ QECC with $d={\rm{min}}~[{\rm{supp}} (\mathcal{S}/\{I\})]$.

A QECC is said to be {\emph{pure}} or non-degenerate if Eq.~(\ref{eq:defEa}) is satisfied with $C(E_a)=0$ for any $E_a$ other than $E_a=I$. Otherwise, the code is called non-pure or  degenerate. Pure codes have the property that every error of support $\left \lfloor (d-1)/2\right \rfloor$ corresponds to a distinct syndrome and allows to {\it identify} its location and its type ($X$, $Y$, or $Z$).   Note that in the literature, error detection is referred to as the possibility to {\emph{know}} that an error occurred, but not to locate nor to identify it. For example the surface code is a $[[2 d^2,2,d]]$ QECC~\cite{criger2016noise}. It allows to  detect $d-1$ errors and correct up to  $\lfloor (d-1)/2 \rfloor$ errors per cycle of stabilizer measurements~\cite{andersen2020repeated}. However it only allows to locate and identify a single 1-qubit error.

According to the above definitions, $m$-uniformity can be related to {\emph{pure}} QECCs: Ref.~\cite{scott2004multipartite} showed that any $m-$uniform state is a pure $[[n,0,m+1]]$ QECC. This result can also be obtained from Sec.~\ref{sec:level3}, together with the statement that any stabilizer state is a QECC with distance $d = \min[{\rm{supp}}( \mathcal{S}/\{I\})]$, and can be generalized to states with a finite number of logical qubits $k$ \cite{huber2020quantum}: If the basis vectors $|i \rangle $ span a $m$-uniform subspace, which is a vector space $\mathcal{V}$ such that each $v \in \mathcal{V}$ is $m-$uniform, then one obtains a pure $[[n,k,m+1]]$ QECC, where $k=\log_2 (\mathrm{dim}(\mathcal{V}))$. Then, using our key result in Sec.~\ref{se:main_result}, we deduce that the $D-$dimensional cluster state is a $[[n,0,2D+1]]$ QECC. For example the 1D cluster state corresponds to a $[[n,0,3]]$ QECC, allowing to locate and identify 1-qubit errors. This case is demonstrated in the next section. The 2D cluster state corresponds to a $[[n,0,5]]$ QECC and allows to detect arbitrary 2-qubit Pauli errors.

\section{Benchmarking errors using the cluster state}
\label{sec:4}
From the above QECC properties, an $m$-uniform state can be used to benchmark errors on a quantum computer that act on at most $\left \lfloor m/2 \right \rfloor $ qubits. Since the $D$-dimensional cluster state is  $2D$-uniform, it allows to detect errors that act on $D$ qubits. We now demonstrate this on the $1D$ cluster state. 
\begin{figure}
\centering
\includegraphics[width = \linewidth]{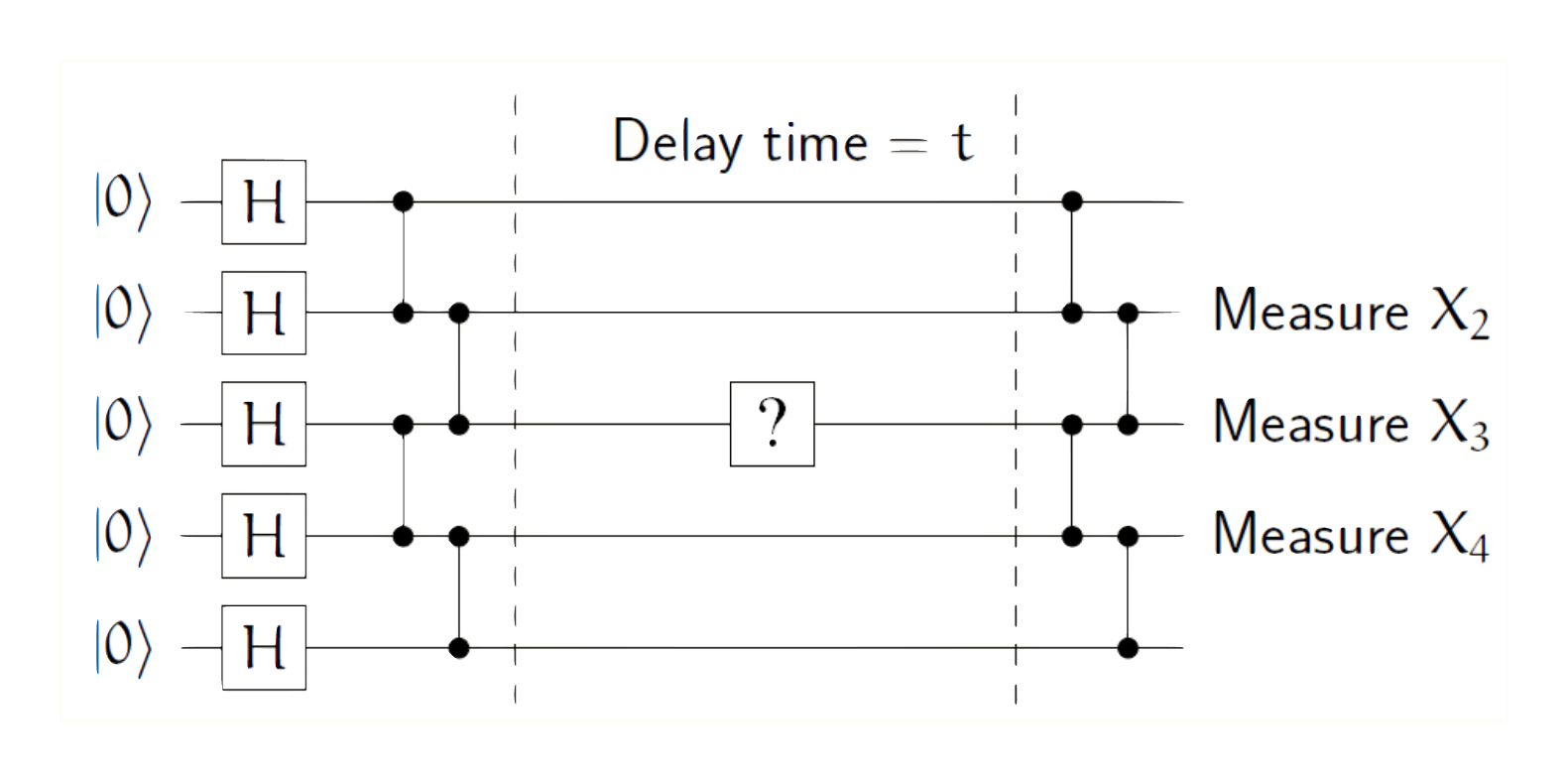}
\caption{Circuit used to benchmark error rates on qubit $3$. \blue{First, a 1D cluster state is prepared with open-boundary conditions on the left. Then, during a delay of time $t$ an error may occur. Finally, the measurements of the stabilizer generators on the right side are used to identify the error. On the real machine errors may occur on any qubit, but dealing with errors on any qubit would require PBCs, i.e. to entangle qubits 1 and 5, which we avoid due to limitations of the actual machine. }}
\label{fig:cluster_error_circuit}
\end{figure}

\begin{figure*}[t]
\begin{tabular}{c c}
(a) Simulation: ZXZ &
(b) Simulation: XZX\\
\includegraphics[width=0.5\linewidth]
{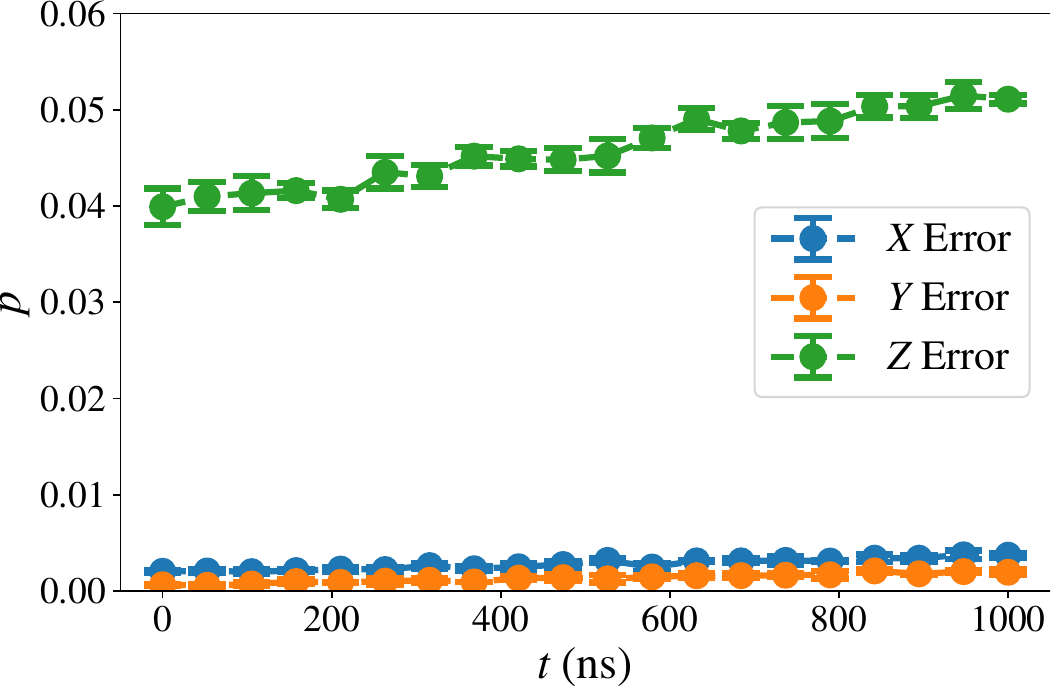} &
\includegraphics[width=0.5\linewidth]
{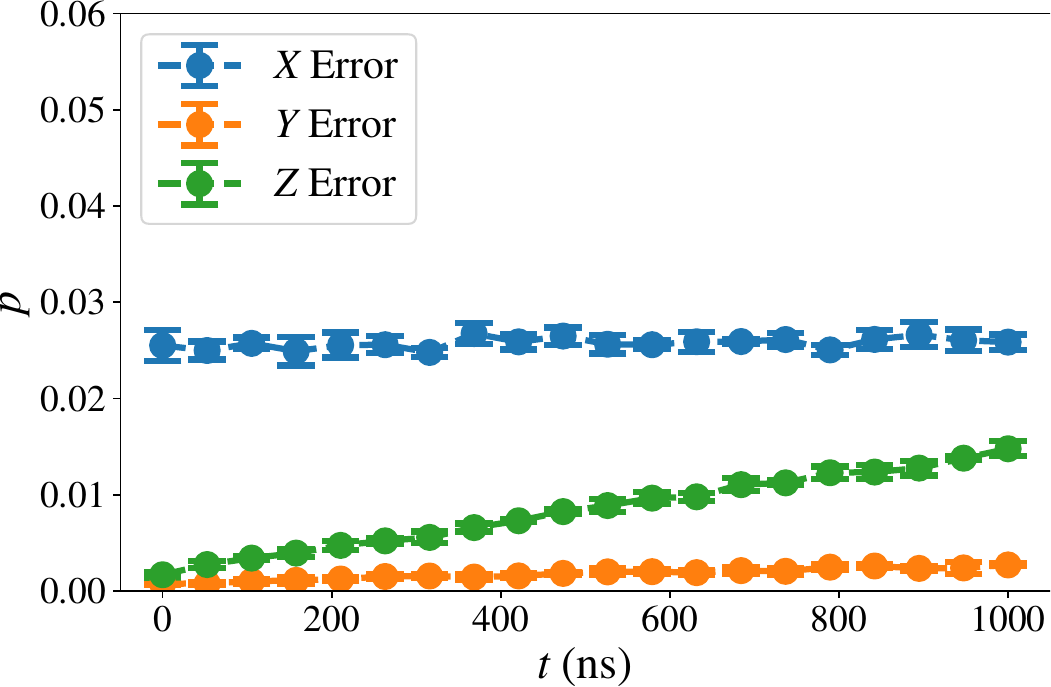}\\
(c) Demonstration: ZXZ &
(d) Demonstration: XZX \\
\includegraphics[width=0.5\linewidth]
{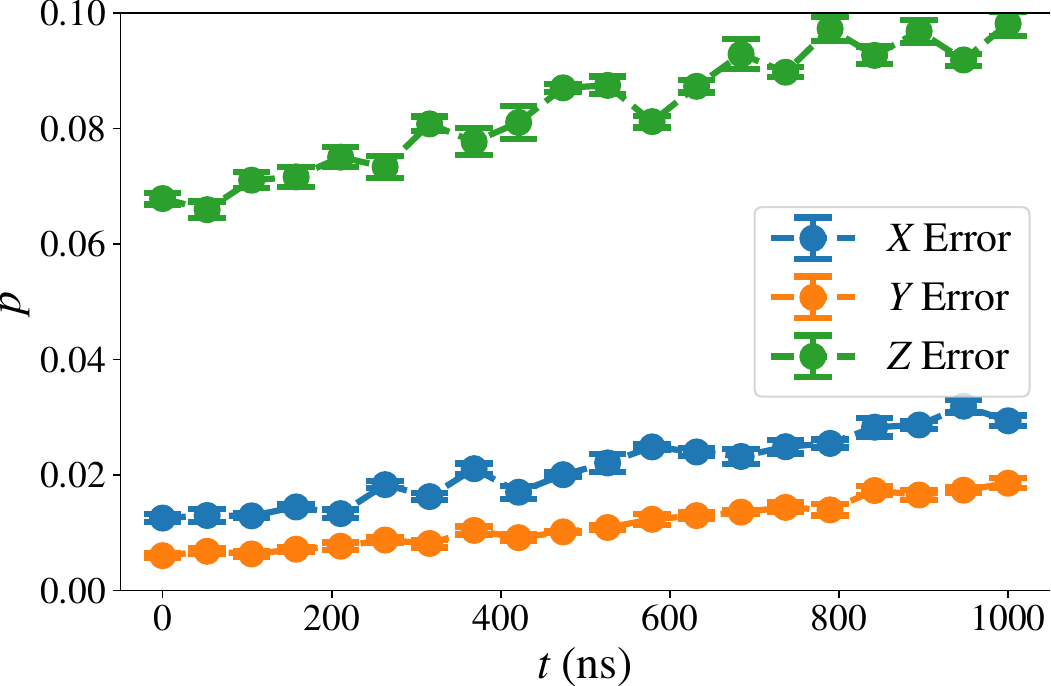}&
\includegraphics[width=0.5\linewidth]
{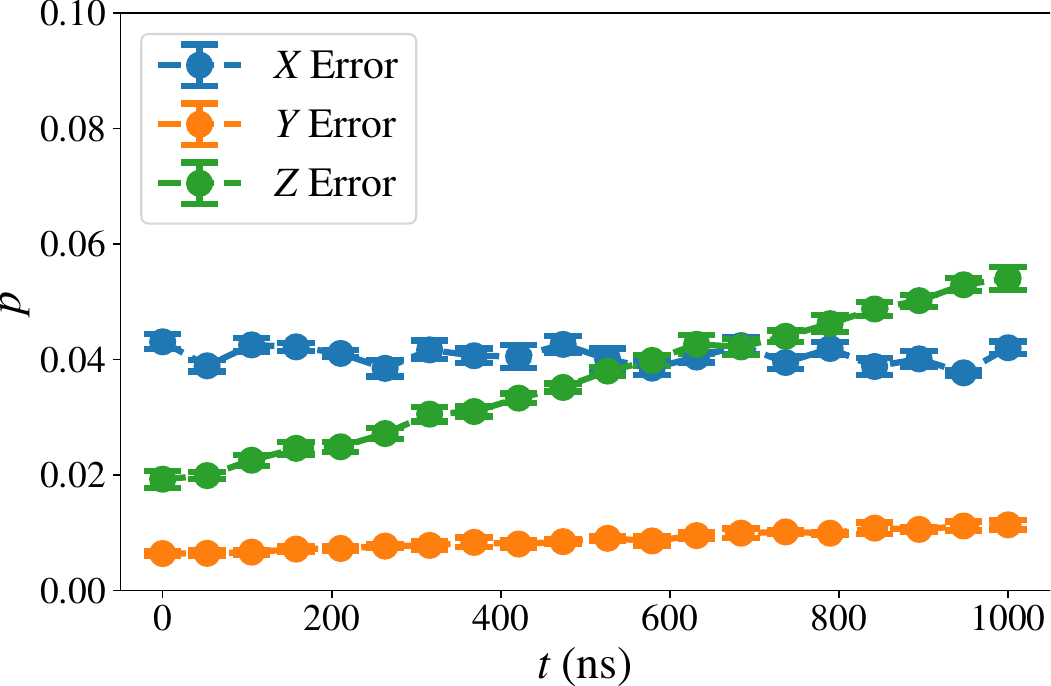}
\end{tabular}
\caption{\label{noise_error} (a) Noisy simulator results of the error benchmarking protocol for the cluster state. The errors are calculated from the measurement results as a function of the delay $t$ as in the main text. Each point is generated from 5 realizations of 20000 shots from which we extract the standard deviation. The slope quantifies the creation of errors by time where the shift from the origin is the result of the measurement errors. (b) The same for the cluster after $X \longleftrightarrow Z$ transformation. During the state preparation (at zero delay) the errors of X and Z are inverted with respect to (a), while during the waiting time (i.e., the slopes) are the same. (c-d) Quantum computer results for the same circuits.}
\end{figure*}

Our noise benchmarking protocol is depicted in Fig.~\ref{fig:cluster_error_circuit}. One first prepares the cluster state from the product state using the unitary transformation $U$ composed of Hadamard (H) and controlled-Z gates (CZ), assuming no errors occur at this stage. During a delay time $t$, an error, or multiple errors, may spontaneously occur, which is depicted as `?' in Fig.~\ref{fig:cluster_error_circuit}. To detect this error, assuming it acted on at most one qubit, one measures the error syndromes through the stabilizers.
Present day quantum computers do not facilitate direct measurement of such stabilizers as they involve simultaneous measurements of multiple qubits. Instead,
the syndromes can be measured by reversing $U$ with $U^\dagger$, assuming perfect fidelity, followed by single qubit measurements as shown in Fig.~\ref{fig:cluster_error_circuit}. By repeating this process many times one obtains a probability distribution of the errors. 

The corresponding error-syndromes detection in a sufficiently long ($n\geq 5$ qubits) $1D$ cluster state with PBC is as follows: if no errors occurred, we always obtain a string of $0$'s. In the case of one $Z_i$ error during the delay, we get $1$ on the $i$-th qubit, instead of $0$. An $X_i$ error evolves to a $101$ pattern on the $i-1,i,i+1$-th qubits. The error $Y_i$ combines the $Z$ and $X$ errors and results in a  $111$ pattern.

In the real machine that we use, the qubits 1 and 5 are physically separated and cannot be entangled directly. Hence, we focus on the OBC cluster Hamiltonian defined by the stabilizers $\{ X_1 Z_2,Z_1X_2Z_3,Z_2X_3Z_4,Z_3X_4Z_5,Z_4 X_5\}$. Then, the $2-$uniformity is spoiled near the edges, and for example, 
 $X_1$ and $Z_2$ errors result in the same syndrome 01000. Thus, we focus on errors that act only on the middle qubit 3. \blue{We added a $?$ mark in Fig.~\ref{fig:cluster_error_circuit} only on qubit 3 because this circuit allows us to deal with errors only at that qubit, despite that in practice errors can occur on any qubit.  Thus,} assuming an error may have occurred only on qubit 3, the error syndromes 00100, 01010 and 01110 allow to determine the probability of $Z$, $X$ and $Y$ errors, respectively. We proceed to apply this protocol on (i) a noisy simulator that mimics the hardware noise, and (ii) a real quantum computer.

Real quantum computers, and noisy simulators that mimic their behaviour, experience both relaxation errors, which change gradually $\ket{1}$ to $\ket{0}$, and dephasing errors, which change gradually $\ket{+}$ to $\ket{-}$ and vice versa. These processes correspond to $X/Y$ and $Z$ errors, respectively, and their characteristic times are commonly denoted by $T_1$ and $T_2$. In superconducting circuits, $T_1$ and $T_2$ are generically of the order of $10-100\mu$sec and satisfy $T_2< T_1$. We first consider a noisy simulator with physical parameters derived from the IBM quantum computer {\it ibmq\_manila} \cite{Qiskit}, see Appendix~\ref{app:q_params}.

Fig.~\ref{noise_error}(a) shows the probability of finding an $X$, $Y$, or $Z$ error in the middle qubit, as a function of the delay time (each point refers to the average over 100,000 shots). We find that the slope, i.e. rate of  $X$ and $Y$ errors is smaller than the slope of  $Z$ errors, in agreement with the expected relation $T_2 < T_1$.

In addition to the slopes, we can see that the curves in Fig.~\ref{noise_error}(a) are shifted differently from the origin. We associate this shift with state-preparation and measurement (SPAM) errors that occur while preparing the cluster state and measuring the stabilizers. These errors occur with a probability that does not depend on the delay time and correspond to a vertical shift of the error curves. Let us denote this readout error as $R_i$ for the i'th qubit  and assume that it has probability $p_i =p \ll 1$. As a result of this error, the pattern 00100 occurs either as a result of a $Z_3$ error or a $R_3$ error. In contrast, $X$ and $Y$ error patterns may be created and equivocally detected as a result of two or three $R_i$ errors, respectively, which have a smaller probability, $p^3 \ll p^2 \ll p$. Thus  $Z$ errors  occur more frequently due to SPAM errors, in agreement with the observed result. To support this error model, we consider a transformed cluster state with $X \leftrightarrow Z $, with stabilizers of the form $XZX$. This state can be prepared by simply applying an additional layer of Hadamard gates before and after the delay. In this case, the error $Z_3$ has syndrome 01010, which can be mistakenly generated by two readout errors.
On the other hand, the $X_3$ error has syndrome $00100$ and can occur due to a single $R_3$ readout error. In Fig.~\ref{noise_error}(b) we present the results from the transformed protocol. When normalizing the readout-errors on the middle qubit (The readout error in Fig.~\ref{noise_error}(a) is approximately twice that in Fig.~\ref{noise_error}(b), see Appendix.~\ref{app:q_params}), we notice that the $X$ error probability is now shifted by approximately the same amount as the $Z$ error of Fig.~\ref{noise_error}(a), and vice-versa. In contrast, we find that the slopes of the errors in the original and transformed circuits are comparable. This finding confirms our hypothesis that the shift of the curves are associated with SPAM errors, while the slopes are due to processes that occur during the waiting time. 

The corresponding results for the real IBM quantum computer {\it ibmq\_manila} are shown in Figs.~\ref{noise_error}(c,d). The observed slopes are higher than in the simulator, indicating that qiskit simulators underestimate the noise in the hardware. For example, the simulator does not take into account the crosstalk between neighboring qubits \cite{Qiskit}, which negatively affects the evolution of entangled states. From the slopes we estimate
\be
T_1 \approx \left( \frac{dp_{X,Y}}{d t} \right)^{-1}\approx 100 {\rm{\mu}sec},~~
T_2 \approx \left( \frac{dp_Z}{d t} \right)^{-1} \approx 30 {\rm{\mu}sec}.
\ee
We emphasize that while typically longitudinal $(T_1)$ and transversal ($T_2$) error rates are measured using separate experiments, our quantum error detection approach based on $m-$uniform states involves a single experiment. Performing this error analysis in longer chains, may allow one to quickly find the qubits with the smallest error rates and improve the computational fidelity. 

\section{$m$-uniform logical spaces}
\label{sec:5}
\begin{figure}[] 
\centering
\includegraphics[width=0.8\linewidth
]{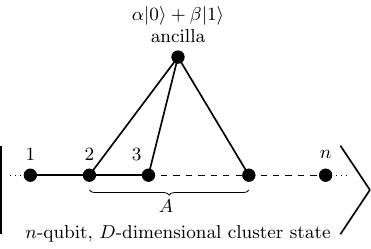}
\caption{To encode information in the cluster state, an ancilla qubit is coupled to a subset of qubits, $A$. The bonds represent controlled-Z gates.}
\label{measurement encoding}
\end{figure}
The cluster state discussed so far does not encode a logical subspace. Is it possible to supplement its error detection ability, with an encoded subspace? Here, we exemplify this possibility via a measurement based protocol~\cite{hein2006entanglement,PhysRevA.65.012308,schlingemann2001stabilizer,Grassl}.

Consider one ancilla qubit prepared in the desired  state $\alpha|0\rangle +\beta |1\rangle$, and a $2D-$uniform state, i.e. the $D$-dimensional cluster state $|cs\rangle$ with $n$ qubits. We couple the ancilla qubit to an arbitrary set of $|A|$ qubits of the cluster state via CZ gates, as shown in Fig.~\ref{measurement encoding}. By measuring the ancilla in the $X$ basis and post-selecting the $X=+1$ outcome, one obtains the state
\begin{equation} \label{encoding with A}
    |\phi\rangle = \alpha|cs\rangle + \beta \prod_{i\in A} Z_i |cs\rangle,
\end{equation}
(see Appendix~\ref{qubit encoding} for a derivation). The state $|\phi\rangle$ encodes one logical qubit. As shown in Appendix~\ref{ minimum central qubit }, the inequality
\begin{equation}
    |A| > 2D(2D+1)
\end{equation}
is a sufficient condition for the encoding in Eq.~(\ref{encoding with A}) to be $2D$-uniform. Thus, if the ancilla qubit is entangled with more than  $2D(2D+1)$ qubits of a $2D$-uniform cluster state, then the resulting logical space is also $2D$-uniform.

\section{Summary}
\label{sec:6}

In this work we explored $m-$uniformity, a measure of multipartite entanglement, in cluster states. $m$-uniform states maximize the entanglement between any $m$ qubits and their surroundings, and can be used for quantum error detection. In contrast to previous studies that focused on quantum states that maximize the uniformity, here we considered the uniformity of cluster states, which are  ground states of local frustration free Hamiltonians and can be realized  on quantum computers with local gates. Our key  result is that $D-$dimensional cluster states are $2D$ uniform.

We introduced a novel application of $m$-uniformity in benchmarking quantum errors. While the amount of uniformity can highly underestimate the support of \emph{correctable} errors, here we emphasized the observation that the uniformity determines the support of \emph{identifiable} errors distinguishing X, Y and Z errors. The $D$-dimensional cluster states allow to detect errors acting independently on $D$ qubits. We demonstrated how the 1D cluster state can be used to benchmark errors on quantum device. This approach allowed us to clearly observe the dominance of one type of errors $(Z)$ over the others ($X$ and $Y$) on the specific machine explored in this work. Applications of the 2D cluster state to benchmark 2-qubit errors and their correlations are left for future study.

An interesting question that deserves further investigation is whether quantum error detection ability extends beyond the special cluster states considered here. A natural candidate for the extension of our work is offered by symmetry protected topological (SPT) states. These states include the cluster states as special cases and were shown to share common properties, for instance as universal resources of measurement-based quantum computations \cite{else2012symmetry,stephen2017computational}. To address the error correction capabilities of SPT states it may be useful to extend the concept of symmetry-resolved entanglement \cite{pollmann2010entanglement,azses2020identification,azses2020symmetry,cornfeld2019entanglement,azses2021observing,fraenkel2020symmetry,azses2022symmetry,ares2022symmetry,monkman2023symmetry} to the multi-partite regime.

\begin{acknowledgments}
We acknowledge the use of IBM Quantum services for this work. The views expressed are those of the authors, and do not reflect the official policy or position of IBM or the IBM Quantum team. SS would like to acknowledge Manisha Goyal, ICTS-TIFR for feedback on some proofs and IISER Kolkata, India, for support in the form of a fellowship. SD would like to acknowledge the MATRICS grant (Grant No. MTR/ 2019/001 043) from the Science and Engineering Research Board (SERB) for funding. We gratefully acknowledge support from the European Research Council (ERC) under the European Unions Horizon 2020 research and innovation programme under grant agreement
No. 951541, ARO (W911NF-20-1-0013) (ES) and the Israel Science Foundation, grant numbers 154/19 (EGDT and ES). We acknowledge enlightening discussions with Robert Raussendorf and Vito Scarola.

S.S. and D.A. contributed equally to this work.
\end{acknowledgments}

\appendix
\section{\blue{Proof of Proposition 1}}\label{se:app_cs_2d_uniform}

We present a proof by contradiction. Consider subsystem $A$ containing  $|A| \le 2D$ qubits. Let us assume that the subgroup $\mathcal{S}_A$ is not the trivial group. That is, there exists at least one matrix $\sigma \neq I$ in $\mathcal{S}_A$. Since $\mathcal{S}_A$ is a subgroup of $\mathcal{S}$, $\sigma$ is the product of  generators of $\mathcal{S}$,
\begin{equation}\label{stabilizer_vertices}
\sigma=s_{\mathbf{v}_1}s_{\mathbf{v}_2}\dots s_{\mathbf{v}_r},
\end{equation}
where $r>0$. The form of the stabilizer generators of the cluster state $s_{\mathbf{v}}$ is such that there is an $X$ acting on a qubit at $\mathbf{v}$. These $X$'s cannot turn into the identity in the product in Eq.~(\ref{stabilizer_vertices}), as the $X$'s for different generators act on different sites. Hence, the set $A$ contains at least $r$ non-identity Pauli operators.

From the set $\big\{\mathbf{v}_1,\mathbf{v}_2,\dots,\mathbf{v}_r\big\}$ there exists one $\mathbf{v}_h$ that has the highest value of the first coordinate. $s_{\mathbf{v}_h}$ is a generator of $\mathcal{S}$ with support on qubits in $\big\{\mathbf{v}_{h},\mathbf{v}_{h}\pm \mathbf{e}_{i}\big\}$, that is, $\mathbf{v}_h$ and its nearest neighbours. As we selected $\mathbf{v}_h$ to have maximal first coordinate, the element $\sigma$ must  contain the Pauli operator $Z_{\mathbf{v}_h +\mathbf{e}_{1}}$  as no other stabilizers from the product can cancel it. Therefore $\mathbf{e}_h +\mathbf{e}_{1}$ should be contained in $A$ as the support of $\sigma$ is in $A$. 

Similarly, from the set $\big\{\mathbf{v}_1,\mathbf{v}_2,\dots,\mathbf{v}_r\big\}$, there exists one generator $\mathbf{v}_l$ whose first coordinate is the minimal one. Therefore, $\mathbf{v}_l -\mathbf{e}_{1}$ should also be contained in $A$.
  
In 1D we conclude that any non-identity $\sigma$ in $A$ must contain at least $r+2$ Pauli's. Continuing this argument to $D$ dimensions, we conclude that ${\rm{supp}}(\sigma) \ge  r+2D$. This contradicts our assumption that $A$ consists of less than $2D+1$ qubits. Therefore, if $|A| \leq 2D$ then $\mathcal{S}_A = \{I\}$.
 $\qed$

\blue{
 \section{Uniformity of different 2D lattices}\label{app:lattice_structures}

\begin{figure}
\begin{tabular}{c c}
(a)  & (b) \\
\includegraphics[width=0.49\linewidth]
{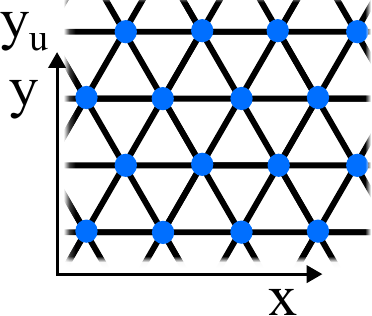} &
\includegraphics[width=0.49\linewidth]
{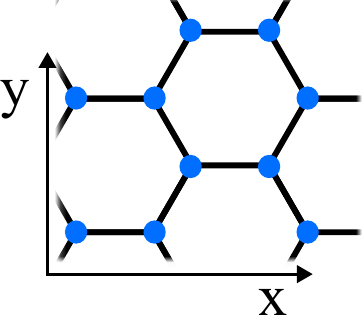}\\
\end{tabular}
\caption{\label{fig_lat} \blue{Vertices represent qubits and  bonds represent controlled-Z gates as in the main text. (a) Triangular lattice. The coordination $y_u$ is the maximal stabilizer generator $y$ coordinate, see text. (b) Hexagonal lattice.}}
\end{figure}

In this appendix we consider the uniformity of different 2D lattices. We focus on the triangular and hexagonal lattices, see Fig.~\ref{fig_lat}(a-b). Here we consider infinite lattices only. Our analysis, which is summarized in Table~\ref{tab:lattice_uniformity}, can be generalized further to different types of lattice structures, such as the 2D Archimedean lattices.

The 2D triangular lattice graph state is generated by stabilizers $s_v$ of $7$ support on any vertex $v$, thus, the uniformity is at most $6$. As in the proof for the 2D cluster state on square lattice, we prove that the support of any stabilizer is at least $7$. Let's focus on stabilizer $\sigma = \prod_v s_v$. First, we notice that each vertex $v$ shares at most 2 neighbors with the vertex $w\neq v$, thus, each additional stabilizer generator at $s_v$ removes at most $2$ support from each other generator. Therefore, as one generator has $7$ support, $2$ generators have at least $2(7-2)=10$ support, and $k$ generators have more than $k(7-2\cdot [k-1])$ support, which is less than $7$ only for $4$ generators and more. Hence, we assume from now on that $\sigma$ has at least $4$ generators, and all left to prove is that $\sigma$ contains additional $3$ $Z$'s to the existing $4$ $X/Y$'s at $v$ for each $s_v$.

As in the proof for the square lattice, we focus on the upper most generators, notating their $y$ coordinate as $y_u$, see Fig.~\ref{fig_lat}(a). If there is only $1$ generator $s_u$ on $y=y_u$, it is clear that $\sigma$ contains two $Z$'s on $y=y_u+1$ induced by $s_u$. If there is more than one such vertex, $\sigma$ still contains $2$ $Z$'s on $y=y_u+1$ from the right most and left most generators at $y=y_u$. Similarly, one has the same analysis from the lower most generators. Therefore, $\sigma$ has at least $4$ $Z$'s as required. Thus, we have proved that the 2D triangular lattice is $6$ uniform.

Let us now focus on the graph state of the 2D hexagonal lattice, see Fig.~\ref{fig_lat}(b). As one generator has now $4$ support, we prove that this lattice is $3$-uniform. Proceeding as in the case of the triangular lattice we focus on $\sigma = \prod_v s_v$. Here, each vertex shares at most $1$ neighbor with any other vertex, thus, $k$ generators have more than $m = k(4-[k-1])$ support, which is less than $4$ only for $5$ generators and more. However, $5$ generators trivially, due to their $X/Y$'s, have more than $4$ support. Hence, the 2D hexagonal lattice graph state is $3$-uniform.

\begin{table}
\caption{\label{tab:lattice_uniformity}\blue{The uniformity of different 2D lattice structures is summarized in this table.}}
\begin{ruledtabular}
\begin{tabular}{cc}
Lattice & uniformity  \\
\hline
Square & 4  \\
Triangular & 6  \\
Hexagonal & 3  \\
\end{tabular}
\end{ruledtabular}
\end{table}
}
\section{\blue{$2D$-uniformity of finite cluster states}}\label{se:proof_finite_lattice}

In this appendix we prove for PBCs that the cluster state with at least 8 vertices in each dimension is $2D-$uniform. 

As in Eq.~(\ref{stabilizer_vertices}), we consider a subsystem $A$ consisting of $2D$ qubits or less, and want to show that no element of the stabilizer group $\mathcal{S}_A$, other than the identity, fits into it. This follows from claim~\ref{eq:stab_local} below. 

First, let us define the ``distance" on the lattice.
\begin{definition}
The Hamming distance between two vertices $\mathbf{v}$ and $\mathbf{w}$, denoted by $|\mathbf{v}-\mathbf{w}|$, is the number of edges in the shortest path on the graph from $\mathbf{v}$ to $\mathbf{w}$. 
\end{definition}
Equivalently, the Hamming distance between two vertices $\mathbf{v}$ and $\mathbf{w}$ is equal to the minimum number of basis vectors ($\mathbf{e_{i}}$) that needs to be added/subtracted to $\mathbf{v}$ to result in $\mathbf{w}$.

\begin{claim} \label{eq:stab_local}
Let $\mathcal{S}_A$ be the stabilizers of the $D$-dimensional cluster state within subsystem $A$ such that $|A|\leq 2D$. If $s_{\mathbf{v}_1}s_{\mathbf{v}_2}s_{\mathbf{v}_3}\dots s_{\mathbf{v}_r}\neq I\in \mathcal{S}_A$, then $|\mathbf{v_i}-\mathbf{v_j}|\leq 4$ for all $\mathbf{v_i},\mathbf{v_j}\in \Big\lbrace \mathbf{v_{1}},\mathbf{v_{2}},\dots,\mathbf{v_{r}}\Big\rbrace $. 
\end{claim}

Before proving claim~\ref{eq:stab_local} by introducing two lemmas, let us draw our main conclusion from it. Consider a specific vertex $\mathbf{v_i}\in \Big\lbrace \mathbf{v_{1}},\mathbf{v_{2}},\dots,\mathbf{v_{r}}\Big\rbrace $. From claim~\ref{eq:stab_local}, all other vertices lie within a ``sphere" of radius $r$, and hence, all of the Pauli operators involved in $s_{\mathbf{v}_1}s_{\mathbf{v}_2}s_{\mathbf{v}_3}\dots s_{\mathbf{v}_r}$ are localized within a sphere of radius $r+1$. Let us use Jung's theorem \cite{jung1901ueber,dekster1985extension}, which relates the diameter of a set to the radius of its bounding sphere, to bound the radius $r$. Jung's theorem states that $r \leq 4 \sqrt{\frac{D}{2(D+1)}}< \sqrt{8}$. Hence, there is a sphere that encloses all the $\mathbf{v_i}$'s with diameter $2r < 2\sqrt{8}<6$, which implies that we have at most 6 vertices in each axis in the sphere. To the sphere diameter we add 2 to cover nearest neighbors, which we notate $d^* = 2(\sqrt{8}+1)<8$, as each stabilizer generator has interaction with only its nearest neighbors. Since the system length $8$ is greater than $d^*$, we can apply the proof by contradiction of the infinite lattice case in Appendix~\ref{se:app_cs_2d_uniform}, since the definition of the ``highest" or ``lowest" value of the $D$ coordinates exists within the sphere.

To prove claim~\ref{eq:stab_local}, we  discuss properties of the stabilizer generators for the $D-$dimensional cluster state. The stabilizer generators  $s_{\mathbf{v}}$ of the cluster state have support over $2D+1$ vertices
 \begin{equation}
\label{eq:stab_gen_locs}
   \Big\lbrace\mathbf{v},\mathbf{v}\pm\mathbf{e_{i}}\Big\rbrace  , 
 \end{equation}
 where $\Big\lbrace\mathbf{v}\pm\mathbf{e_{i}}\Big\rbrace$ is the neighbourhood of $\mathbf{v}$. The product of stabilizer generators $s_{\mathbf{v}}$ and $s_{\mathbf{w}}$ does not have support over the intersection of their neighbourhoods due to cancellations of $Z$'s. We show below how two stabilizer generators can, at most, intersect at two qubits.
 \begin{lemma}
 \label{lem:lem1}
 The neighborhoods of two stabilizer generators $s_{\mathbf{v}}$ and $s_{\mathbf{w}}$ overlap at most in two vertices for lattices with PBC where all axes are of length $\geq 5$.
 \end{lemma}

\textbf{Proof:}
Eq.~(\ref{eq:stab_gen_locs}) implies that $s_{\mathbf{v}}$ has support in a ball of radius 1 around $\mathbf{v}$. Therefore, $s_{\mathbf{v}}$ and $s_{\mathbf{w}}$ intersect only if $|\mathbf{v}-\mathbf{w}|\leq 2$. Let us check the intersection case by case.
\begin{itemize}
    \item $|\mathbf{v}-\mathbf{w}|=1$:   
    Then there exists an $\mathbf{e_k}$, such that
    \begin{equation}
        \mathbf{v} + \mathbf{e_{k}}=\mathbf{w}.
    \end{equation}
    The distance between neighbourhood points of $\mathbf{v}$ and $\mathbf{w}$ is
    \begin{equation}
        |\mathbf{v}\pm\mathbf{e_{i}}-\mathbf{w}\mp \mathbf{e_{j}}|=|\mathbf{e_{i}}\mp \mathbf{e_{j}}-\mathbf{e_{k}}|\geq 1.
    \end{equation}
    This implies that thee neighbourhoods of $\mathbf{v}$ and $\mathbf{w}$ are disjoint.
    \item $|\mathbf{v}-\mathbf{w}|=2$:     
    Then 
    \begin{equation}
\mathbf{v}+\mathbf{e_{k_1}} + \mathbf{e_{k_2}}=\mathbf{w},
    \end{equation}
    for some indices $k_1,k_2$. Let us check the intersection case by case.
        \item $k_1=k_2$:
        \begin{equation}
            \mathbf{v}+2\mathbf{e_{k_1}}=\mathbf{w}\implies\mathbf{v}+\mathbf{e_{k_1}}=\mathbf{w}-\mathbf{e_{k_1}}.
        \end{equation}
        The neighbourhoods intersect at one point $\mathbf{v}+\mathbf{e_{k_1}}$ only.
    \item $k_1 \neq k_2$:
    \begin{equation}
        \mathbf{v}+\mathbf{e_{k_1}} + \mathbf{e_{k_2}}=\mathbf{w}
    \end{equation}
    \begin{equation}
        \implies \mathbf{v} + \mathbf{e_{k_1}}=\mathbf{w} - \mathbf{e_{k_2}}
    \end{equation}
    and additionally
    \begin{equation}
        \implies \mathbf{v}+\mathbf{e_{k_2}}=\mathbf{w}-\mathbf{e_{k_1}}.
    \end{equation}
    Therefore, the neighbourhoods intersect at two points: $\mathbf{v}+\mathbf{e_{k_1}}$ and $\mathbf{v}+\mathbf{e_{k_2}}$. \qed
\end{itemize}
The fact that the neighbourhoods of two stabilizer generators intersect at most at two vertices implies the following lemma
\begin{lemma}\label{lemma 2}
Let $\sigma=s_{\mathbf{v}_1}s_{\mathbf{v}_2}\dots s_{\mathbf{v_r}} \in\mathcal{S}_A$. Then, for any $s_{\mathbf{v}_i}$ there is a set of $D+1$ points in $A$ that are localized around $\mathbf{v}_i$ with at most $2$ distance.
\end{lemma}
\textbf{Proof:} 
Let us write $\sigma=s_{\mathbf{v}_1}s_{\mathbf{v}_2}\dots s_{\mathbf{v_r}} \in\mathcal{S}_A$.
Consider $s_{\mathbf{v}_k}\in\Big\{s_{\mathbf{v}_1},s_{\mathbf{v}_2},s_{\mathbf{v}_3},\dots,s_{\mathbf{v}_r}\Big\}$. $s_{\mathbf{v}_k}$ has support over $2D$ neighbouring qubits of $\mathbf{v}_k$, which we notate $\mathcal{N}(\mathbf{v}_k)$. As of Lemma.~\ref{lem:lem1}, any other stabilizer generator in $\Big\{s_{\mathbf{v}_1},s_{\mathbf{v}_2},s_{\mathbf{v}_3},\dots,s_{\mathbf{v}_r}\Big\}/\Big\{s_{\mathbf{v}_k}\Big\}$ intersects at most with two vertices in $\mathcal{N}(\mathbf{v}_k)$. Then, for $v_j \in \mathcal{N}(\mathbf{v}_k)$, either $v_j \in A$ or it is in the intersection of $s_{\mathbf{v}_k}$  and another stabilizer generator. As each intersection contains at most 2 points in $\mathcal{N}(\mathbf{v}_k)$, the minimal number of points $\mathbf{w} \in A$ that are localized within distance 2 around $\mathbf{v}_k$ such that $|\mathbf{w}-\mathbf{v}_k| \leq 2$ is $D+1$ (including $\mathbf{v}_k$ itself). 

\textbf{Proof of Claim 1:} Consider a $D$-dimensional cluster Hamiltonian with PBC and a subsystem $A$. Let us prove claim 1 by contradiction. Assume that $s_{\mathbf{v}_1}s_{\mathbf{v}_2}s_{\mathbf{v}_3}\dots s_{\mathbf{v}_r}\neq I\in \mathcal{S}_{A}$ and $|A|\leq 2D$. For any $s_{\mathbf{v_i}}$ in $\{s_{\mathbf{v}_1},s_{\mathbf{v}_2},s_{\mathbf{v}_3},\dots,s_{\mathbf{v}_r} \}$ we get at least $D$ vertices in $A$ that are at most 2-distance away from $\mathbf{v_{i}}$ as of Lemma~\ref{lemma 2}. Choosing $\mathbf{v_{i}}$ and $\mathbf{v_{j}}$ such that $|\mathbf{v_{i}}-\mathbf{v_{j}}|>4$, then the $D$ extra points from $\mathbf{v_{i}}$ and $\mathbf{v_{j}}$ cannot overlap and hence the total number of points in $A$ becomes greater than $2D$. $\qed$

\begin{figure}[t] 
\centering
\includegraphics[width=\linewidth]{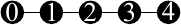}
\caption{\label{fig:ibmq_scheme} Qubit scheme of {\it ibmq\_manila}.}
\end{figure}

\section{\blue{Specifications of  the quantum computer {\it ibmq\_manila}}}\label{app:q_params}

In this appendix we provide the different parameters of the quantum computer {\it ibmq\_manila}, see scheme in Fig.~\ref{fig:ibmq_scheme}, at the running time of the quantum circuits and their noisy simulations. The noise parameters are extracted from the last calibration before the circuits run, see Tables~\ref{tab:qubit_params_zxz} and \ref{tab:qubit_params_xzx}. The noisy simulations were done using the usual Qiskit software package with the standard noise model, see List~\ref{alg:noise_calibration} and Ref.~\cite{noisemodelqiskit}. To get the full calibration properties at the time of running the circuits, see List~\ref{alg:noise_calibration}.

\onecolumngrid

\pythonstyle
\begin{lstlisting}[label = {alg:noise_calibration}, caption = {Python Code}]
# Qiskit version: {'qiskit-terra': '0.23.3', 'qiskit-aer': '0.12.0', 'qiskit-ignis': '0.6.0', 'qiskit-ibmq-provider': '0.20.2', 'qiskit': '0.42.1', 'qiskit-nature': None, 'qiskit-finance': None, 'qiskit-optimization': None, 'qiskit-machine-learning': None}

# Extracting calibration at the time of the quantum demonstration
calib_props = IBMQ.load_account().get_backend('ibmq_manila').properties(datetime=insert_datetime)

# Initiating noise model
NoiseModel.from_backend_properties(calib_props)
\end{lstlisting}

\twocolumngrid

\begin{table*}
\caption{\label{tab:qubit_params_zxz} Here we provide all the properties (as of its last calibration with respect to the circuit run) of the IBM quantum computer {\it ibmq\_manila} which we have used to get the results for the quantum demonstration in Fig.~\ref{noise_error}(c) at Jan 12, 2023 12:13 PM Pacific standard time}
\begin{ruledtabular}
\begin{tabular}{ccccccccc}
Qubit name & Frequency [GHz] & T1 [us] & T2 [us] & Readout error & ID error & $\sqrt{X}$ error & Pauli-X error & CNOT error \\
Q0 & 4.9623 & 187.8607 & 99.2866 & 4.71e-02 & 2.1453e-04 & 2.1453e-04 & 2.1453e-04 & 6.5306e-03 \\
Q1 & 4.8379 & 150.3745 & 73.712 & 1.89e-02 & 2.2034e-04 & 2.2034e-04 & 2.2034e-04 & [6.5306e-03, 8.5347e-03] \\
Q2 & 5.0373 & 154.846 & 25.6775 & 3.35e-02 & 2.5892e-04 & 2.5892e-04 & 2.5892e-04 & [8.5347e-03, 6.8509e-03] \\
Q3 & 4.951 & 193.3112 & 63.4785 & 2.52e-02 & 2.0920e-04 & 2.0920e-04 & 2.0920e-04 & [6.8509e-03, 7.1415e-03] \\
Q4 & 5.0651 & 156.7893 & 40.7793 & 3.39e-02 & 6.3812e-04 & 6.3812e-04 & 6.3812e-04 & 7.1415e-03 \\
\end{tabular}
\end{ruledtabular}
\end{table*}

\begin{table*}
\caption{\label{tab:qubit_params_xzx} Here we provide all the properties (as of its last calibration with respect to the circuit run) of the IBM quantum computer {\it ibmq\_manila} which we have used to get the results for the quantum demonstration in Fig.~\ref{noise_error}(d) at Jan 14, 2023 5:23 PM Pacific standard time}
\begin{ruledtabular}
\begin{tabular}{ccccccccc}
Qubit name & Frequency [GHz] & T1 [us] & T2 [us] & Readout error & ID error & $\sqrt{X}$ error & Pauli-X error & CNOT error \\
Q0 & 4.9623 & 82.21 & 115.5179 & 2.20e-02 & 1.8141e-04 & 1.8141e-04 & 1.8141e-04 & 6.2379e-03 \\
Q1 & 4.8379 & 172.9769 & 72.0529 & 3.05e-02 & 2.6893e-04 & 2.6893e-04 & 2.6893e-04 & [6.2379e-03, 9.5845e-03] \\
Q2 & 5.0372 & 125.2094 & 28.3401 & 1.73e-02 & 2.2990e-04 & 2.2990e-04 & 2.2990e-04 & [9.5845e-03, 6.4435e-03] \\
Q3 & 4.951 & 156.7664 & 56.1308 & 3.12e-02 & 2.0040e-04 & 2.0040e-04 & 2.0040e-04 & [6.4435e-03, 6.9814e-03] \\
Q4 & 5.0651 & 141.3944 & 40.0803 & 2.98e-02 & 7.2360e-04 & 7.2360e-04 & 7.2360e-04 & 6.9814e-03 \\
\end{tabular}
\end{ruledtabular}
\end{table*}

\section{\blue{Derivation of Eq.~(\ref{encoding with A})}}\label{qubit encoding}

In this appendix we derive Eq.~(\ref{encoding with A}). Connecting the $i$-th qubit of the cluster state $|cs\rangle$ with the ancilla ($n+1$) qubit yields
\begin{equation} \label{1-edge}
\mathrm{CZ}_{i,n+1}|cs\rangle(\alpha|0\rangle+\beta|1\rangle) ,
\end{equation}
where $\mathrm{CZ}_{i,n+1}= (1/2)(I+Z_i + Z_{n+1}-Z_i Z_{n+1}) $ is the controlled-Z gate acting on the $i$-th and $n+1$-th qubits.
Therefore the expression in Eq.~(\ref{1-edge}) can be expanded as
\begin{multline}
\frac{1}{2}(I+Z_i + Z_{n+1}-Z_i Z_{n+1})(\alpha |cs\rangle|0\rangle + \beta |cs\rangle|1\rangle) 
\\
=\alpha|cs\rangle|0\rangle+\beta Z_i |cs\rangle|1\rangle \\= \frac{1}{\sqrt{2}}\bigg((\alpha|cs\rangle+\beta Z_i |cs\rangle)|+\rangle+(\alpha|cs\rangle-\beta Z_i |cs\rangle ) |-\rangle\bigg) .
\end{multline}
Measuring the $n+1$-th qubit in the $X$ basis and post selecting the $X=+1$ eigenvector yields the encoded state
\begin{equation}
  \alpha|cs\rangle+\beta Z_i|cs\rangle.
\end{equation}
This procedure can be generalized to the case where the ancilla qubit is connected to an arbitrary set $A$ of qubits, leading to Eq.~(\ref{encoding with A}).

\section{\blue{2D-uniform cluster state with a logical subspace}} \label{ minimum central qubit }
In this appendix we derive a lower bound for the number of qubits $A$ of the cluster state that the central qubit needs to be entangled with in order for the resulting logical space to retain the $2D$-uniformity of the cluster state. 

We first denote the encoded state in Eq.~(\ref{encoding with A}) as
\begin{equation}
    |\phi\rangle = \alpha |cs\rangle + \beta \mathcal{Z}_A | cs\rangle
\end{equation}
where, $\mathcal{Z}_A = \prod_{i\in A} Z_i $. Now, $|\phi\rangle$ is $m$-uniform iff $\langle\phi|\hat{O}(m)|\phi\rangle = 0$ for $\hat{O}(m)$ being a string of up to $m$ Pauli matrices acting non-trivially  in $A$. We will show that $m=2D$.

This leads to the requirement
\begin{multline}\label{four terms}
 0=   \langle \phi |\hat{O}(m)|\phi\rangle = |\alpha|^2 \langle cs|\hat{O}(m)| cs\rangle +\alpha^*\beta\langle cs|\hat{O}(m)\mathcal{Z}_A| cs\rangle \\+\alpha\beta^*\langle cs|\mathcal{Z}_A\hat{O}(m)| cs\rangle +|\beta|^2\langle cs|\mathcal{Z}_A\hat{O}(m)\mathcal{Z}_A| cs\rangle . 
\end{multline}
Since $|cs\rangle$ is $2D$-uniform and the operator $\mathcal{Z}_A\hat{O}(m)\mathcal{Z}_A$ acts at most on $m$ qubits, the first and fourth terms of the right hand side of Eq.~(\ref{four terms}) vanish for $m=2D$. 

We assume that $|A|>m$. The second term, $\langle cs|\hat{O}(m)\mathcal{Z}_A| cs\rangle=0$ if $\hat{O}(m)\mathcal{Z}_A$ is not contained in the generalized Bloch expansion \footnote{The generalized Bloch expansion of a pure $n$ qubit state $|\psi\rangle$ is the expansion of $|\psi\rangle\langle\psi|$ in the basis of tensor products of Pauli matrices} of $| cs\rangle$. Since $|cs\rangle$ is a stabilizer state, all of its Bloch expansion terms are generated by the stabilizers $\{s_{\mathbf{v}}\}$. Now, $\hat{O}(m)$ contains a maximum of $m$-$X$'s and/or $Y$'s. Therefore $\hat{O}(m)\mathcal{Z}_A$ is a product of at most $m$ stabilizer generators $s_i$'s. Since each stabilizer generator contributes $2D$ $Z$'s, the product of $m$ such stabilizer generators is an operator with $Z$'s acting on a maximum of $2Dm$ qubits. Also, $\hat{O}(m)\mathcal{Z}_A$ contains a minimum of $|A|-m$ $Z$'s. Therefore, a sufficient condition for $\hat{O}(m)\mathcal{Z}_A$ to be excluded from the Bloch expansion of $|\phi\rangle$ is for the maximum number of qubits that can be acted upon by $Z$'s coming from the stabilizer generators to be lesser than the minimum number of $Z$'s possible in $\hat{O}(m)\mathcal{Z}_A$. This yields
\begin{equation}
    2Dm<|A|-m\implies m(2D+1)<|A|.
\end{equation}
Substituting $m=2D$ we obtain the anticipated condition
\begin{equation}\label{central qubit inequality}
    2D(2D+1)<|A|.
\end{equation}
The same argument applies for the third term in Eq.~(\ref{four terms}). In 1D, this lower bound gives $|A| \ge 7$.

\end{document}